\documentclass[preprint, 1p]{elsarticle}

\usepackage[english]{babel}
\usepackage{lipsum}
\usepackage[english]{babel}
\usepackage{natbib}
\usepackage[colorlinks=true,linkcolor=black, citecolor=blue, urlcolor=blue]{hyperref}
\usepackage{graphicx}
\usepackage{placeins}
\usepackage{amsmath}
\usepackage{amssymb}
\usepackage{tikz}
\usetikzlibrary{arrows,chains,calc,shapes,automata,positioning,decorations.markings}
\usepackage{pgfplots}
\usepackage{color,soul}

\begin{document}

\begin{frontmatter}
	
\title{Physics-informed deep learning for flow and deformation in poroelastic media}
\author[]{Yared W. Bekele}
\address{{\scriptsize Rock and Soil Mechanics Group, SINTEF AS, Trondheim, Norway}}
\ead{yared.bekele@sintef.no}

\begin{abstract}
A physics-informed neural network is presented for poroelastic problems with coupled flow and deformation processes. The governing equilibrium and mass balance equations are discussed and specific derivations for two-dimensional cases are presented. A fully-connected deep neural network is used for training. Barry and Mercer's source problem with time-dependent fluid injection/extraction in an idealized poroelastic medium, which has an exact analytical solution, is used as a numerical example. A random sample from the analytical solution is used as training data and the performance of the model is tested by predicting the solution on the entire domain after training. The deep learning model predicts the horizontal and vertical deformations well while the error in the predicted pore pressure predictions is slightly higher because of the sparsity of the pore pressure values.      
\end{abstract}


\end{frontmatter}

\section{Introduction}

\noindent In recent years, physics-informed neural networks (PINNs) have created a new trend at the intersection of machine learning and computational modeling research. Such models involve physical governing equations as constraints in the neural network such that training is performed both on example data and the governing equations. In addition to PINNs, various names are used by different researchers to refer to the concept and the most common ones include \emph{physics-based deep learning}, \emph{theory-guided data science} and \emph{deep hidden physics models}. In general, the aims of these applications include improving the efficiency, accuracy and generalization capability of numerical methods for the solution of PDEs.

Since the pioneering work on PINNs by \citet{raissi2019physics}, where well-known partial differential equations (PDEs) such as Burgers' equation and Schroedinger's equation are investigated, the concept has been applied to various problems in computational science and engineering. Application to quantitative finance and statistical mechanics  was presented by \citet{al2018solving} where the Black-Scholes and Focker-Planck PDEs, respectively, are applied as physical constraints.   The application areas are increasing rapidly with different variations in the general methodology. A deep learning-based solution of the Euler equations for modeling high speed flows was presented by \citet{mao2020physics} where physics-informed neural networks were used for forward and inverse problems. Deep learning for computational fluid dynamics, in particular for vortex-induced vibrations, was presented by \citet{raissi2019deep}. A related work for predictive large-eddy-simulation wall modeling was presented by \citet{yang2019predictive}. The solution of time-dependent stochastic PDEs using physics-informed neural networks by learning in the modal space was demonstrated by \citet{zhang2019learning}. Application of deep learning, with physics-informed recurrent neural networks, to fleet prognosis was presented by \citet{nascimento2019fleet}. Bending analysis of Kirchhoff plates using a deep learning approach was shown by \citet{guo2019deep}. Deep learning-based study of linear and non-linear diffusion equations to learn parameters and unknown constitutive relationships was presented by \citet{tartakovsky2018learning}. Other recent and related studies are those by \citet{zhang2019quantifying}, \citet{yang2019adversarial}, \citet{meng2020composite}, \citet{sun2020surrogate}, \citet{huang2019predictive}, \citet{tipireddy2019comparative}, \citet{jia2020physics}, \citet{zheng2019physics}, \citet{xu2020physics}, \citet{sahli2020physics}, \citet{zhang2020physics}, \citet{kadeethum2020physics} and \citet{fraces2020physics}.

In this paper, application of PINNs to problems of poroelasticity is presented. The governing equations of poroelasticity describe the coupled flow and deformation processes in porous media. Barry and Mercer's poroelastic problem with an 'exact' analytical solution is used to train a deep neural network model where the poroelastic governing equations are applied as constraints. In the following, the coupled governing PDEs of poroelasticity are first described. The architecture of the deep learning model is then described. A numerical example is then presented to show the performance of the deep learning model in comparison with the existing analytical solution.

\section{Governing Equations of Poroelasticity}

The governing equations of poroelasticity are a combination of the overall mass balance equation, the equilibrium or linear momentum balance equation and the linear elastic constitutive equations for stress-strain relationships.

\subsection{Mass balance equation}

The general mass balance equation for a two-phase porous medium (fluid saturated porous solid matrix) under isothermal conditions, obtained from superposition of the individual phase mass balance equations, is given by
\begin{equation}
\alpha \nabla \cdot \dot{\boldsymbol{\overline{u}}} + \left( \frac{\alpha-n}{K_\mathrm{s}} + \frac{n}{K_\mathrm{f}} \right) \frac{\partial \bar{p}}{\partial \bar{t}} + \nabla \cdot \boldsymbol{\overline{w}} = Q, \label{eq:genmasbal} 
\end{equation}
where $ \alpha = 1 - K/K_\mathrm{s} $ is Biot's coefficient, $ \boldsymbol{\overline{u}} $ is the solid deformation vector, $ n $ is the porosity, $ K_\mathrm{s} $ is bulk modulus of the solid, $ K_\mathrm{f} $ is the bulk modulus of the fluid, $ K $ is total bulk modulus of the porous medium, $ \bar{p} $ is the pore fluid pressure, $ \boldsymbol{\overline{w}} $ is the fluid velocity vector and $ Q $ is a fluid source or sink term. For a porous medium with incompressible constituents, i.e. $ 1/K_\mathrm{s} = 1/K_\mathrm{f} = 0 $, the mass balance equation reduces to
\begin{equation}
\nabla \cdot \dot{\boldsymbol{\overline{u}}} + \nabla \cdot \boldsymbol{\overline{w}} = Q. \label{eq:masbalincompressible} 
\end{equation}
The fluid velocity as described by Darcy's law, assuming flow in the porous medium is driven by pressure gradients only, can be expressed as
\begin{equation}
\boldsymbol{\overline{w}} = -\frac{\mathbf k}{\gamma_\mathrm{f}} \nabla \bar{p},
\label{eq:darcyslaw}
\end{equation}
where $ \mathbf k $ is the hydraulic conductivity matrix and $ \gamma_\mathrm{f} $ is the unit weight of the fluid. For an isotropic porous medium, the hydraulic conductivity is the same in all directions of flow and the magnitude of the hydraulic conductivity, $ k $, replaces the hydraulic conductivity matrix in the equation above. For a two-dimensional problem, the deformation vector is $ \boldsymbol{\overline{u}} = \left\lbrace \bar{u}, \bar{v} \right\rbrace^\intercal $ where $\bar{u}$ and $\bar{v}$ are the deformations along the $x$ and $z$ directions, respectively. Introducing the deformation vector and combining equations \eqref{eq:genmasbal} and \eqref{eq:darcyslaw} gives
\begin{equation}
\frac{\partial}{\partial \bar{t}} \left( \frac{\partial \bar{u}}{\partial \bar{x}} + \frac{\partial \bar{v}}{\partial \bar{z}} \right)  - \frac{k}{\gamma_\mathrm{f}} \left( \frac{\partial^2 \bar{p}}{\partial \bar{x}^2} + \frac{\partial^2 \bar{p}}{\partial \bar{z}^2} \right) = Q
\label{eq:masbalxz}
\end{equation}
wherein the hydraulic conductivity is assumed to be isotropic as described earlier.

\subsection{Linear momentum balance equation}

The overall linear momentum balance or equilibrium equation for a two-phase porous medium is given by
\begin{equation}
\nabla \cdot \boldsymbol{\overline{\sigma}} + \rho \boldsymbol{\overline{b}} = \mathbf 0,
\label{eq:genmombal}
\end{equation}
where $ \boldsymbol{\overline{\sigma}} $ is the total stress carried by both the solid and fluid constituents of the porous medium, $ \rho $ is the overall density of the porous medium and $ \boldsymbol{\overline{b}} $ represents body forces. The effective stress concept differentiates the stresses carried by the solid and fluid constituents by introducing an effective stress $ \boldsymbol{\overline{\sigma}}^\prime = \boldsymbol{\overline{\sigma}} - \bar{p} \boldsymbol{I} $, where $ \boldsymbol{I} $ is an identity matrix. Introducing this into the previous equation and disregarding body forces gives
\begin{equation}
\nabla \cdot \boldsymbol{\overline{\sigma}}^\prime + \nabla \bar{p} = \mathbf 0.
\label{eq:mombaleff}
\end{equation}
The equations of linear elasticity relate the effective stress $ \boldsymbol{\overline{\sigma}}^\prime $ and the deformation of the porous medium through
\begin{equation}
\boldsymbol{\overline{\sigma}}^\prime = 2 \mu \boldsymbol{\overline{\varepsilon}} + \lambda \text{tr}(\boldsymbol{\overline{\varepsilon}}) \mathbf I,
\label{eq:elasticity}
\end{equation}
where $ \mu $ and $ \lambda $ are the Lam\'e parameters and $ \boldsymbol{\overline{\varepsilon}} $ is the infinitesimal strain tensor. The infinitesimal strain tensor is related to deformation through
\begin{equation}
\boldsymbol{\overline{\varepsilon}} = \frac{1}{2} \left( \nabla \boldsymbol{\overline{u}} + (\nabla \boldsymbol{\overline{u}})^\intercal \right).
\label{eq:strain}
\end{equation}
For a two-dimensional problem, combining equations \eqref{eq:mombaleff}, \eqref{eq:elasticity} and \eqref{eq:strain} results in the following two equilibrium equations for an isotropic poroelastic medium:
\begin{align}
(\lambda + 2 \mu) \frac{\partial^2 \bar{u}}{\partial \bar{x}^2} + \mu \frac{\partial^2 \bar{u}}{\partial \bar{z}^2} + (\lambda + \mu) \frac{\partial^2 \bar{v}}{\partial \bar{x} \partial \bar{z}} + \frac{\partial \bar{p}}{\partial \bar{x}} &= 0, \label{eq:mombalx} \\
\mu \frac{\partial^2 \bar{v}}{\partial \bar{x}^2} + (\lambda + 2 \mu) \frac{\partial^2 \bar{v}}{\partial \bar{z}^2} + (\lambda + \mu) \frac{\partial^2 \bar{u}}{\partial \bar{x} \partial \bar{z}} + \frac{\partial \bar{p}}{\partial \bar{z}} &= 0. \label{eq:mombalz} 
\end{align}

\subsection{Nondimensionalized poroelastic equations}

Equations \eqref{eq:masbalxz}, \eqref{eq:mombalx} and \eqref{eq:mombalz} are the the governing equations of poroelasticity in two dimensions where the field variables are the deformations in the $x$ and $z$ directions, $\bar{u}$ and $\bar{v}$, and the pore fluid pressure $ \bar{p} $. These equations can be nondimensionalized with respect to a certain dimension of the porous medium and the material parameters; \citet{barry1999exact}. Nondimensionalizing with respect to a typical length $l$ of the porous medium, the Lam\'e parameters $\lambda$ and $\mu$, the hydraulic conductivity $k$ and the fluid unit weight $\gamma_\mathrm{f}$ implies
\begin{align}
x &= \frac{\bar{x}}{l}, \quad z = \frac{\bar{z}}{l}, \quad t = \frac{(\lambda + 2\mu)k}{\gamma_\mathrm{f} l^2} \bar{t}, \label{eq:nondimxzt} \\
u &= \frac{\bar{u}}{l}, \quad v = \frac{\bar{v}}{l}, \quad p = \frac{\bar{p}}{\lambda + 2\mu}. \label{eq:nondimuvp}  
\end{align}
Applying equations \eqref{eq:nondimxzt} and \eqref{eq:nondimuvp} to equations \eqref{eq:masbalxz}, \eqref{eq:mombalx} and \eqref{eq:mombalz} results in the following nondimensional poroelastic governing equations:
\begin{align}
\frac{\partial^2 u}{\partial t \partial x} + \frac{\partial^2 v}{\partial t \partial z} - \frac{\partial^2 p}{\partial x^2} - \frac{\partial^2 p}{\partial z^2} - \beta Q &= 0, \label{eq:masbalnondim} \\
(\eta+1) \frac{\partial^2 u}{\partial x^2} + \frac{\partial^2 u}{\partial z^2} + \eta \frac{\partial^2 v}{\partial x \partial z} + (\eta+1) \frac{\partial p}{\partial x} &= 0, \label{eq:mombalnondimx} \\ 
\frac{\partial^2 v}{\partial x^2} + (\eta+1) \frac{\partial^2 v}{\partial z^2} + \eta \frac{\partial^2 u}{\partial x \partial z} + (\eta+1) \frac{\partial p}{\partial z} &= 0, \label{eq:mombalnondimz}    
\end{align}
where $\eta$ and $\beta$ are nondimensional parameters which are functions of the real material parameters and are given by
\begin{equation}
\eta = 1 + \frac{\lambda}{\mu} \quad \text{and} \quad \beta = \frac{\gamma_\mathrm{f} h^2}{(\lambda + 2\mu) k}.
\end{equation}

\section{Deep Learning Model}

The architecture of the deep learning model, the model hyper-parameters, the model performance metrics for training and automatic differentiation, which enables applying PDEs as constraints, are discussed in this section. 

\subsection{Neural network architecture}

\begin{figure}[h]
	\centering
	\includegraphics[scale=0.33]{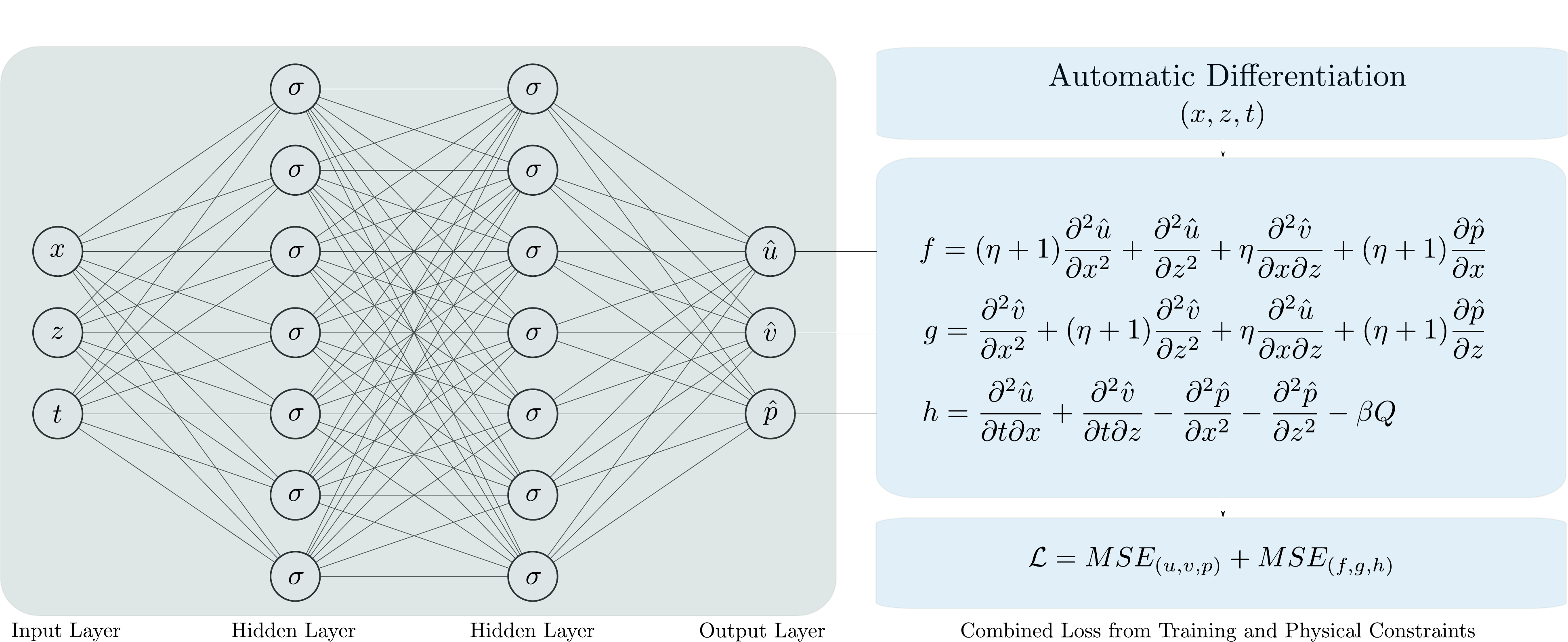}
	\caption{Illustration of the neural network architecture with input, hidden and output layers. The activation function used at the hidden units is $\sigma(\boldmath x) = ReLU(\boldmath x)$. Automatic differentiation is used to compute the partial derivatives in the governing equations, which are then used as physical constraints to optimize together with the prediction error based on training data. The number of hidden layers and hidden units in this figure is for illustration only; the actual number of layers and hidden units used are discussed in the numerical example section.}
	\label{fig:nn}
\end{figure}

The deep learning model used for training is a fully-connected neural network where the number of hidden layers and hidden units per layer is adjusted depending on the problem under consideration. The input layer of the neural network uses sampled data from the spatial and temporal bounds of the poroelastic problem i.e. $\left\lbrace x, z, t\right\rbrace $ as training data. The neural network is design to predict the deformation of the porous medium and the pore fluid pressure i.e. $\left\lbrace \hat{u}, \hat{v}, \hat{p} \right\rbrace $, which is compared with the sample training data of $\left\lbrace u, v, p \right\rbrace $ to measure the performance of the model. The predicted deformations and pore pressure values are used to compute the residuals of the governing mass balance and equilibrium equations such that the model is trained to minimize the computed residuals.  An illustration of the neural network architecture is shown in Figure~\ref{fig:nn}. 

\subsection{Automatic differentiation}

The poroelastic governing equations are applied as constraints in the neural network model by using automatic differentiation (AD) to evaluate the terms in the PDEs based on the model predicted outputs. This makes AD an in important element of the deep learning model. It should be emphasized that AD is different from other methods of computing derivatives using computers; \citet{baydin2017automatic}. The four ways of computing derivatives on computers are: a) manually obtaining the derivatives and coding them; b) numerical differentiation using finite difference approximations; c) computer-based symbolic differentiation and subsequent evaluation based on the algebraic expression; and d) AD, which is the enabler in the neural network model here. The main difference between AD and the other methods is that AD computes the numerical values of the derivatives by using the rules of symbolic differentiation but by keeping track of the derivative values at different stages of numerical operation instead of obtaining the final expressions for the derivatives. This is done by exploiting the fact that any derivative computation, no matter how complex, is composed of a sequence of elementary arithmetic operations and elementary function evaluations. It applies the chain rule repeatedly to these operations until the desired derivative is computed. The fact that AD keeps track of the derivative values makes it computationally superior to the other two commonly used methods of computing derivatives, namely numerical differentiation and symbolic differentiation. The approach used by AD also makes it accurate at machine precision levels. For our problem here, once a deep neural network is designed with the input and output layers described in the previous section, AD is used to estimate the derivative terms in the governing equilibrium and mass balance equations. The model implementation is performed in \verb|TensorFlow| and its AD capability is utilized. \verb|TensorFlow| is an open-source software developed by the Google Brain team at Google and it is a symbolic math library that can be used for different tasks such as data flow, differentiable programming and machine learning; see \citet{abadi2016tensorflow}.

\subsection{Model training and hyper-parameters} 

The training data involves a selected sample from an exact analytical solution where data from the model spatial and temporal discretization bounds $\left\lbrace x, z, t \right\rbrace $ are used as inputs to the neural network and the analytical deformation and pore pressure values $\left\lbrace u, v, p \right\rbrace $ are used as training outputs. The performance of the model is measured by comparing the model-predicted deformation and pore pressure values with the training data based on a performance metrics. The metric chosen here is the mean squared error and the \emph{training loss} corresponding to each field variable is computed from
\begin{align}
	MSE_u &= \frac{1}{N} \sum_{k=1}^{N} \left| u(x_k,z_k,t_k) - \hat{u}(x_k,z_k,t_k) \right|^2 \\
	MSE_v &= \frac{1}{N} \sum_{k=1}^{N} \left| v(x_k,z_k,t_k) - \hat{v}(x_k,z_k,t_k) \right|^2 \\
	MSE_p &= \frac{1}{N} \sum_{k=1}^{N} \left| p(x_k,z_k,t_k) - \hat{p}(x_k,z_k,t_k) \right|^2 
\end{align}
where $N$ represents the number of training data points. The \emph{training loss} is calculated as the sum of the training losses for each field variable i.e.
\begin{equation}
	MSE_t = MSE_u + MSE_v + MSE_p.
\end{equation}
The physical constraint based on the governing PDEs is applied by first evaluating the derivatives of the model predicted outputs at the input training points. The residuals corresponding to each field variable are defined and computed from
\begin{align}
	f(x,z,t) &= (\eta+1) \dfrac{\partial^2 \hat{u}}{\partial x^2} + \dfrac{\partial^2 \hat{u}}{\partial z^2} + \eta \dfrac{\partial^2 \hat{v}}{\partial x \partial z} + (\eta+1) \dfrac{\partial \hat{p}}{\partial x} \\ 
	g(x,z,t) &= \dfrac{\partial^2 \hat{v}}{\partial x^2} + (\eta+1) \dfrac{\partial^2 \hat{v}}{\partial z^2} + \eta \dfrac{\partial^2 \hat{u}}{\partial x \partial z} + (\eta+1) \dfrac{\partial \hat{p}}{\partial z} \\
	h(x,z,t) &= \dfrac{\partial^2 \hat{u}}{\partial t \partial x} + \dfrac{\partial^2 \hat{v}}{\partial t \partial z} - \dfrac{\partial^2 \hat{p}}{\partial x^2} - \dfrac{\partial^2 \hat{p}}{\partial z^2} - \beta Q	
	\label{eq:fgh}
\end{align}
The constraint loss corresponding to each variable is the calculated from
\begin{align}
	MSE_f &= \frac{1}{N} \sum_{k=1}^{N} \left| f(x_k,z_k,t_k) \right|^2 \\
	MSE_g &= \frac{1}{N} \sum_{k=1}^{N} \left| g(x_k,z_k,t_k) \right|^2 \\
	MSE_h &= \frac{1}{N} \sum_{k=1}^{N} \left| h(x_k,z_k,t_k) \right|^2 
\end{align}
The total \emph{constraint loss} is then obtained from
\begin{equation}
	MSE_c = MSE_f + MSE_g + MSE_h.
\end{equation}
The final \emph{total loss} is then computed as the sum of the training loss and the constraint loss i.e.
\begin{equation}
	\mathcal{L} = MSE_t + MSE_c.
\end{equation}
The model training is performed such that the total loss $\mathcal{L}$ is minimized by the model optimizer. The loss minimization is performed by making the necessary adjustments to the hyper-parameters for \emph{bias-variance} tradeoff. The model hyper-parameters tuned during training include the \emph{number of layers}, \emph{number of hidden units}, \emph{batch size} and \emph{learning rate}. If the model is observed to have high bias, the neural network architecture (number of layers and hidden units) are adjusted and/or the model is trained longer. In case of high variance, the amount of training data is increased and/or the neural network architecture is adjusted. The batch size is adjusted to control the number of samples from the training data that are passed into the model before updating the trainable model parameters. The total loss here is minimized using the \emph{Adam optimizer} where its associated learning rate is tuned during the training process.       

\FloatBarrier

\section{Numerical Example}

\subsection{The Problem of Barry and Mercer}

The source problem of Barry and Mercer involves the deformation of a poroelastic medium due to injection and extraction from a point source in the medium; \citet{barry1999exact}. An exact analytical solution is obtained for the problem by choosing the boundary conditions carefully. The idealization of the boundary conditions makes the problem not entirely realistic but still resembles an oil injection/extraction problem; \citet{phillips2005finite}.

\begin{figure}[h]
	\centering
	\includegraphics[scale=0.7]{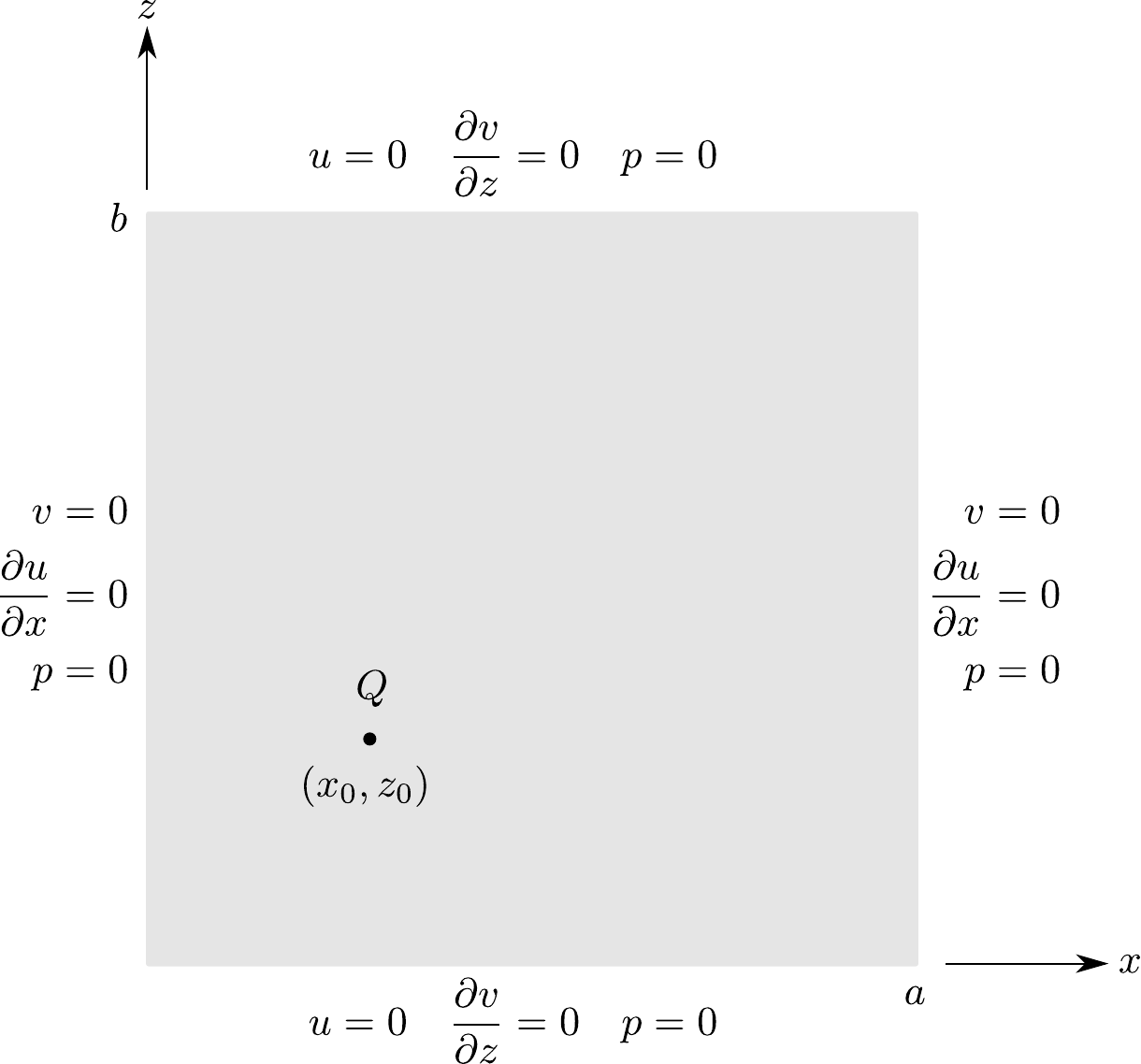}
	\caption{Illustration of the boundary conditions for the Barry and Mercer problem. The dimensions of the poroelastic medium are $x=a$ and $z=b$, which are nondimensionalized corresponding to the nondimensionalized governing equations. A point source Q is applied at an arbitrary location $(x_0, z_0)$. }
	\label{fig:barry_and_mercer}
\end{figure}

An illustration of the problem is shown in Figure~\ref{fig:barry_and_mercer}. The poroelastic domain has dimensions of $x=a$ and $z=b$. We consider the source problem variant from Barry and Mercer's original treatise here where all the four boundaries are assumed to be drained i.e. $p=0$ all four boundaries. The displacement boundary conditions are chosen such that $u=0$ and $\frac{\partial v}{\partial z}=0$ along the boundaries $z=0$ and $z=b$ and $v=0$ and $\frac{\partial u}{\partial x}=0$ along the boundaries $x=0$ and $x=a$. An oscillating point source at $(x_0, z_0)$ is applied and is given by
\begin{equation}
	Q(x,z,t) = \delta(x-x_0) \delta(z-z_0) \sin \omega t
\end{equation}
where $\delta$ represents the Dirac delta function and $\omega$ is the frequency of the oscillation.   

The analytical solutions are obtained by performing sine and cosine transformations, where the transformed variables are defined as $(n,q,s)$, producing a transformed system of equations. Inverse Laplace transformation of the solution in terms of $(n,q,s)$ results in the the following expressions:
\begin{align}
	\hat{p}(n,q,t) &= -\dfrac{\beta \sin \lambda_n x_0 \sin \lambda_q z_0}{\lambda_{nq}^2 + \omega^2} \left( \lambda_{nq} \sin \omega t - \omega \cos \omega t + \omega e^{-\lambda_{nq} t} \right) \\
	\hat{u}(n,q,t) &= \dfrac{\lambda_n}{\lambda_{nq}} \hat{p}(n,q,t) \\
	\hat{v}(n,q,t) &= \dfrac{\lambda_q}{\lambda_{nq}} \hat{p}(n,q,t).  
\end{align}
where
\begin{equation}
	\lambda_n = \frac{n\pi}{a}, \qquad  \lambda_q = \frac{q\pi}{b} \qquad \text{and} \qquad \lambda_{nq} = \lambda_n^2 + \lambda_q^2.
\end{equation}
The final solutions for the deformations and pore pressure, after performing sine and cosine transformations, are given by 
\begin{align}
	u(x,z,t) &= \dfrac{4}{ab} \sum_{n=1}^{\infty} \sum_{q=1}^{\infty} \hat{u}(n,q,t) \cos \lambda_n x \sin \lambda_q z \\
	v(x,z,t) &= \dfrac{4}{ab} \sum_{n=1}^{\infty} \sum_{q=1}^{\infty} \hat{v}(n,q,t) \sin \lambda_n x \cos \lambda_q z \\
	p(x,z,t) &= \dfrac{4}{ab} \sum_{n=1}^{\infty} \sum_{q=1}^{\infty} \hat{p}(n,q,t) \sin \lambda_n x \sin \lambda_q z.
 \end{align} 

\subsection{Results}

An analytical solution for Barry and Mercer's source problem is generated with the following parameters: $a=1, b=1, \beta=2, \omega=1, x_0=0.25, z_0=0.25$ and $\eta=1.5$. The boundary conditions are as illustrated in Figure~\ref{fig:barry_and_mercer}. The spatial discretization grid size used is $N_x \times N_z=41 \times 41$ with the time bounds of $\left[ 0, 2\pi \right] $ where the time steps are defined with a temporal discretization size of $N_t=41$. This solution generates $N_x \times N_y \times N_t$ data points with known inputs $\left\lbrace x, z, t \right\rbrace $ and known outputs $\left\lbrace u, v, p \right\rbrace $.

\begin{figure}[h]
	\centering
	\includegraphics[scale=0.4]{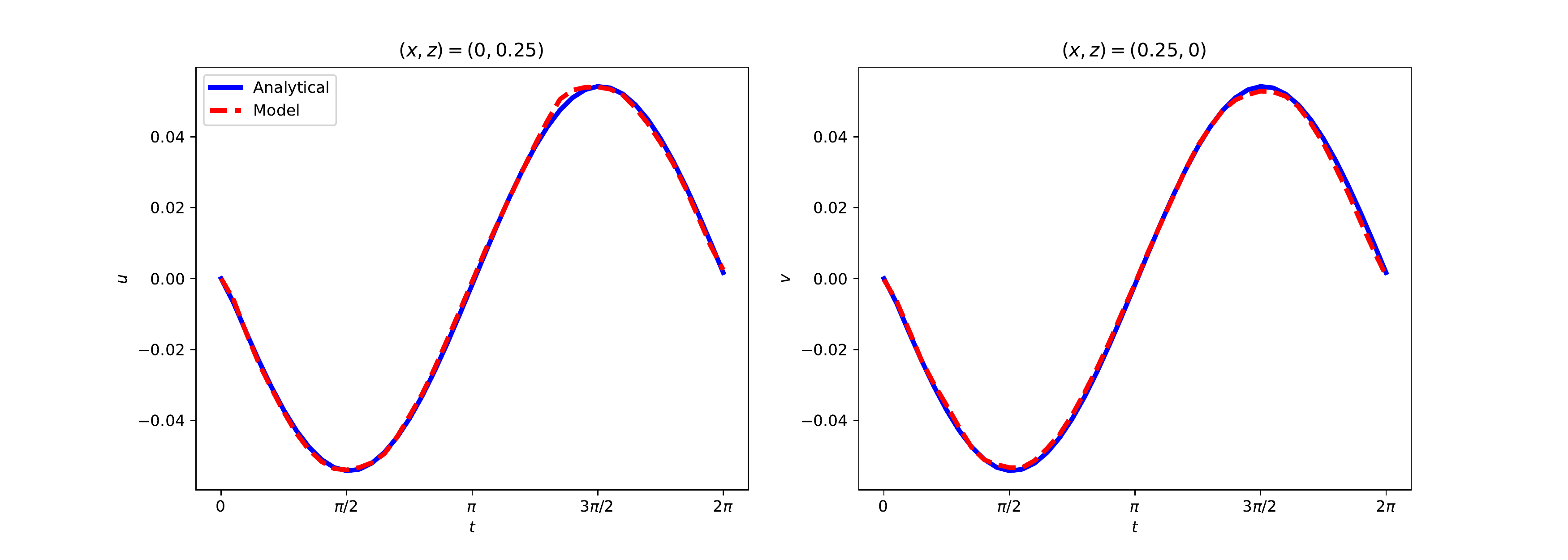}
	\caption{Time evolution of the horizontal displacement $u$ of point $(x,z)=(0,0.25)$ (left) and the vertical displacement $v$ of the point $(x,z)=(0.25,0)$ (right).}
	\label{fig:uv_vs_t}
\end{figure}

A neural network with 5 hidden layers and 40 hidden units at each layer is set up as the model to be trained, after a certain bias-variance tradeoff adjustments. From the available known inputs and outputs in the analytical solution, a training data of size $N=20000$ is randomly selected. A batch size of 5000 is used for training. Adam optimizer is used for training with a learning rate of 0.001. 
\begin{figure}[h]
	\centering
	\includegraphics[scale=0.45]{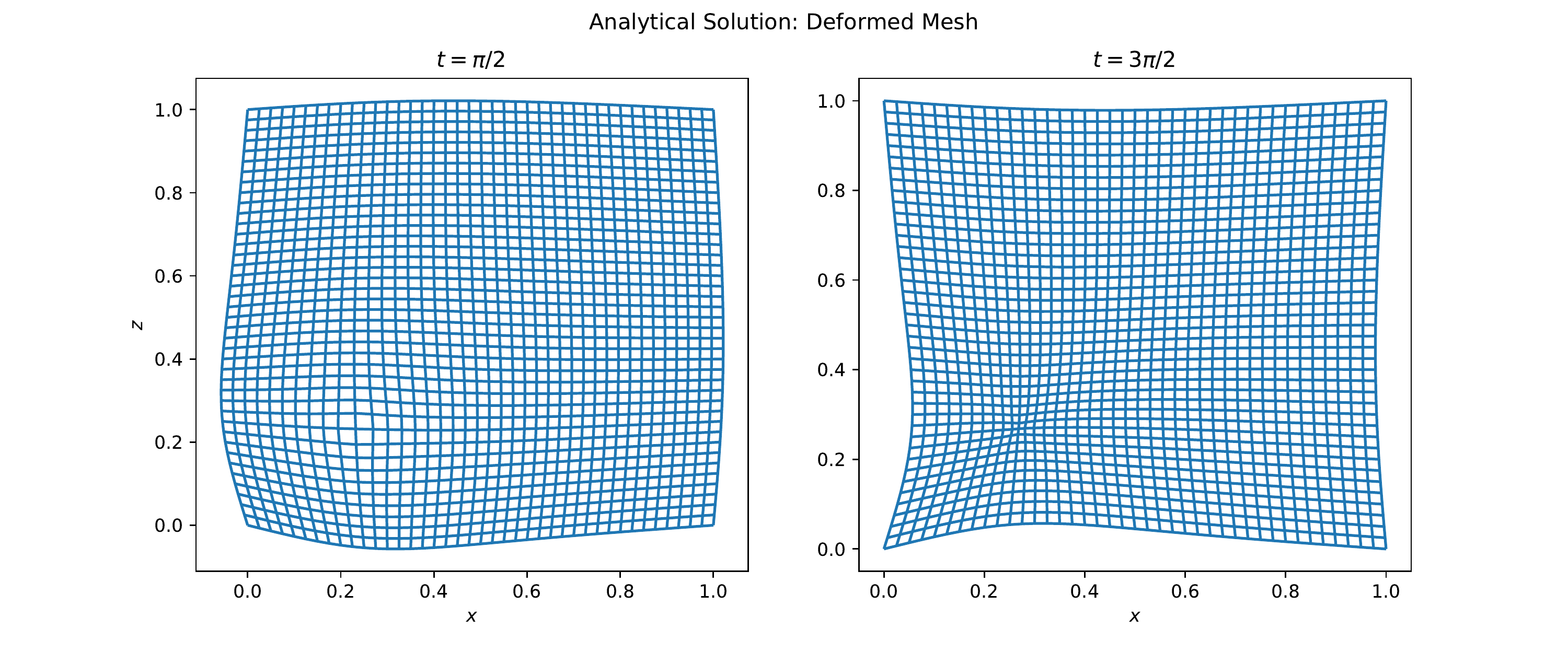}
	\includegraphics[scale=0.45]{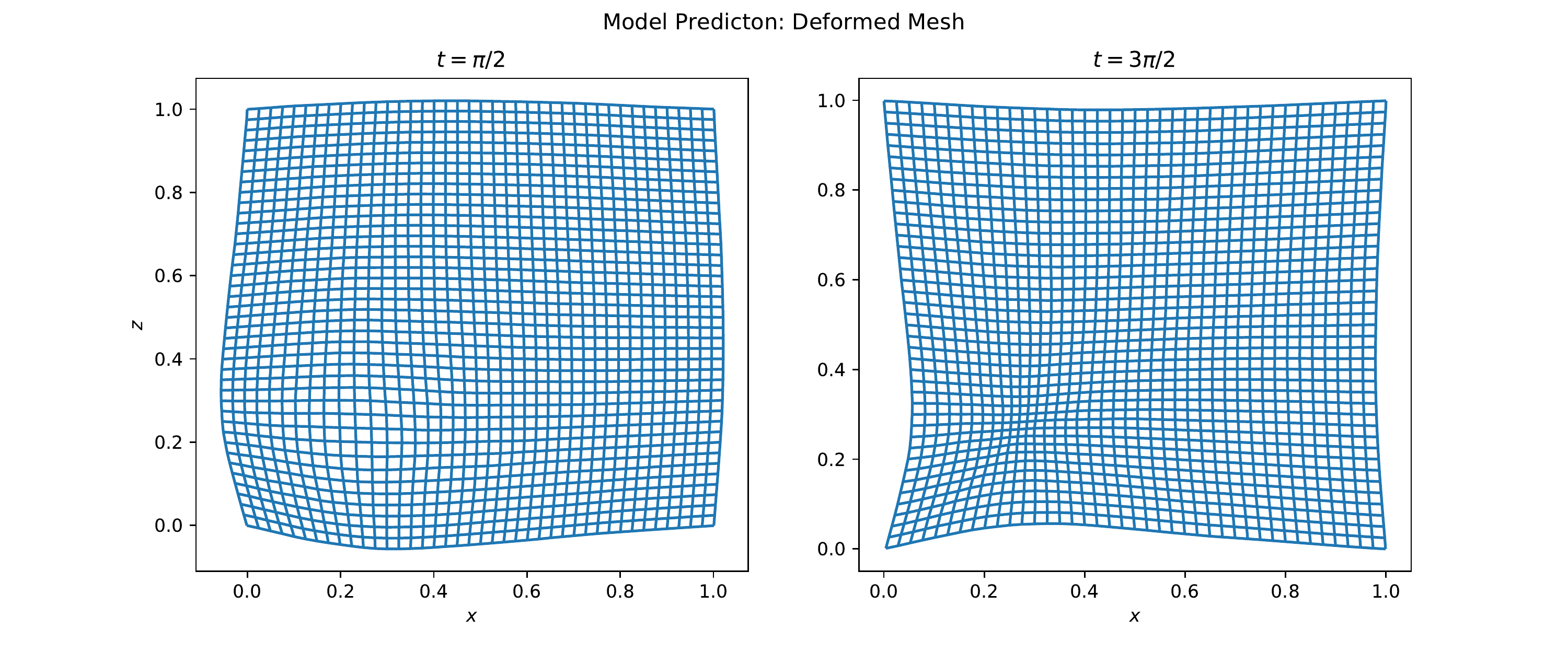}
	\caption{Deformation of the porous medium from the analytical solution (top row) and the deep learning model prediction (bottom row) as a result of injection and extraction at the point $(x_0,z_0)=(0.25,0.25)$ according to the source function $Q(x,z,t) = \delta(x-x_0) \delta(z-z_0) \sin \omega t$. The parameters used are $a=1, b=1, \beta=2, \alpha=1, $ and $\eta=1.5$.}
	\label{fig:anasol_model_deformed_mesh}
\end{figure}
The training data inputs $\left\lbrace x, z, t \right\rbrace $ are normalized to be between 0 and 1 before being passed into the neural network. Similarly, the known outputs $\left\lbrace u, v, p \right\rbrace $ are normalized to be between -1 and 1 before being used to measure the model performance against the model predictions $\left\lbrace \hat{u}, \hat{v}, \hat{p} \right\rbrace $. The final model predictions are inverted to their original scales once they are found satisfactory i.e. when the training is completed. The training is performed for 5000 epochs. The trained neural network is used to predict the solution over the whole domain. The deformed meshes of the porous medium at time steps $t=\frac{\pi}{2}$ and $t=\frac{3\pi}{2}$ are shown in Figure~\ref{fig:anasol_model_deformed_mesh} for both the analytical solution and the neural network prediction. It is noticed that the deep learning model captures the physics of the problem and produces the expected deformations in the porous medium. A closer comparison of the the deformations for selected points is shown in Figure~\ref{fig:uv_vs_t} where the plot on the left shows the horizontal displacement $u$ of the point $(x,z)=(0,0.25)$ and the plot on the right shows the vertical displacement $v$ of the point $(x,z)=(0.25,0)$, both versus time. It can be seen that the neural network predicts the evolution of the deformations well.     

\begin{figure}[h]
	\centering
	\includegraphics[scale=0.4]{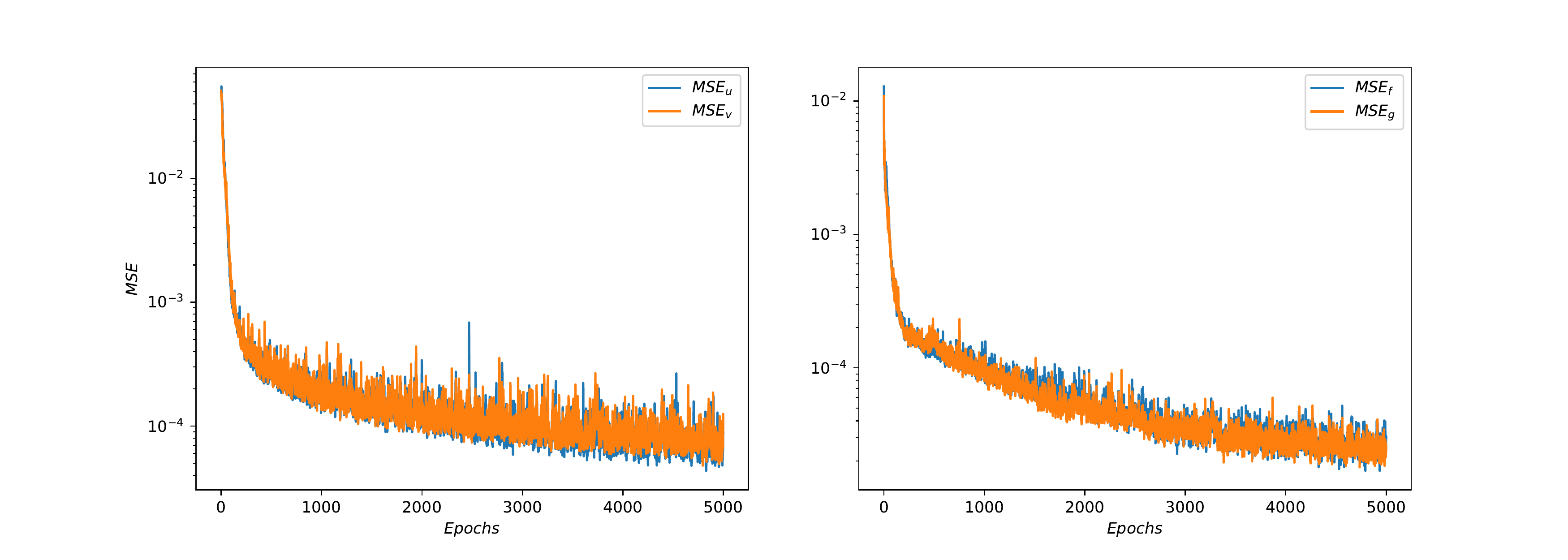}
	\caption{Evolution of the mean squared errors $MSE_u$ and $MSE_v$ (left) and $MSE_f$ and $MSE_g$ (right) as the training progresses i.e. as a function of the number of epochs.}
	\label{fig:MSEs_vs_Epochs}
\end{figure}

A look at the performance metrics of the deep learning model shows that the neural network is trained with a good bias-variance tradeoff. This can be seen in the development of the mean squared errors shown in Figure~\ref{fig:MSEs_vs_Epochs}, corresponding to $MSE_u$ and $MSE_v$ on the left plot and $MSE_f$ and $MSE_g$ on the right plot. The $ L_2 $ norms of the errors for the predicted horizontal and vertical displacements were found to be $3.826 \times 10^{-2}$ and $5.263 \times 10^{-2}$, respectively. 

As an injection and extraction problem in a poroelastic medium, the pore pressure values are concentrated at the injection/extraction point. This makes the magnitude of the pore pressures zero or close to zero over the majority of the domain. Hence, the neural network set up used here is not best suited to train on such a sparse data. This is reflected by the $ L_2 $ norm of the error in the predicted pore pressure, which is found to be $8.071 \times 10^{-2}$. The predicted pore pressure values, however, did not have a significant effect on the horizontal and vertical deformations when used in the constraining differential equations. For the specific problem studied here, if higher accuracy is required for the predicted pore pressures, it may be necessary to use a deep learning model with dual neural networks or data processing techniques best suited to process the training pore pressure data. The model presented here works reasonably well as a demonstration of constraining neural networks using the governing equations of poroelasticity.     

\FloatBarrier

\section{Summary and Conclusions}

A physics-informed deep neural network is presented for flow and deformation in poroelastic media. The governing differential equations for poroelasticity are the mass balance and equilibrium equations complemented respectively by Darcy's law and an elastic stress-strain relationship. An isotropic incompressible porous medium is assumed here in the derivation of the mass balance equation. The equilibrium equations are derived for static poroelastic problems where infinitesimal strains are assumed. Lam\'e's parameters are used in defining the elastic stress-strain relationship. Both governing equations are specifically derived for a two dimensional case with the demonstration problem here in mind. The governing equations in 2D are nondimensionalized with respect to a chosen set parameters. 

The architecture of the neural network is discussed, where a fully-connected network is used here. The neural network inputs are selected data from the spatial and temporal domains of the poroelastic problem and the outputs are the deformations (horizontal and vertical) and the pore pressure. The governing partial differential equations are applied as constraints in the neural network by evaluating their residuals based on the predicted deformations and pore pressure values. This is enabled by automatic differentiation, a discussion for which is briefly presented. The model performance is measured by using mean squared errors as a metric. The mean squared errors for the deformations and for the pore pressure are evaluated based on the model predictions and the known training data. The training loss is defined as the sum of the mean squared errors for each field variable. In a similar way, mean squared errors are defined for the residuals of the constraining partial differential equations. The constraint loss is defined as the sum of the mean squared errors for each governing equation. The final total loss is defined as the sum of the training and constraint losses.

Barry and Mercer's source problem with time-dependent fluid injection and extraction in an idealized poroelastic medium is used as a demonstration problem. The problem has an exact analytical solution which is used to generate the training input and output data. Training of the network is performed by selecting a random sample from the analytical solution. The hyper-parameters of the deep learning model are tuned during training to optimize the performance of the model with respect to both bias and variance. Once the training is completed, the model is used to predict the solution over the entire domain of the problem. The results show that the trained model predicts the horizontal and vertical deformations with a reasonably good accuracy. The error in the predicted pore pressures was found to be slightly higher due to the sparse nature of the pore pressure distribution in the domain. But the higher errors in the predicted pore pressures did not have a significant effect on the final predicted deformations when used in the constraining differential equations. A higher accuracy for the predicted pore pressures may be achieved by tweaking the neural network architecture or by applying advanced data processing techniques best suited to a highly sparse dataset.     

\bibliographystyle{elsarticle-num-names}
\bibliography{pinn_poro}

\end{document}